\documentstyle[aps,multicol]{revtex}
\def\ket#1{| #1\rangle}
\begin{document}

\title{Quantum nonlocality, Bell inequalities and the memory loophole}
\author{ Jonathan Barrett$^{1}$, Daniel Collins$^{2,3}$, Lucien Hardy$^{4}$, 
Adrian Kent$^{3,5}$ and Sandu Popescu$^{2,3}$}
\address{$^1$ DAMTP, Centre for Mathematical Sciences, University of Cambridge,
Wilberforce Road, Cambridge, CB3 OWA, U.K.\\  
 $^2$ H.H. Wills Physics Laboratory, University of Bristol, Tyndall
Avenue, Bristol, BS8 1TL, U.K.\\
$^3$ Hewlett-Packard Laboratories, Filton Road, Stoke Gifford, 
Bristol, BS34 8QZ, U.K.\\
$^{4}$ Centre for Quantum Computation, Clarendon Laboratory, Parks Road,
Oxford, OX1 3PU, U.K.\\
$^{5}$ on leave from $1$}
\date{April 2002; revised May 2002}

\maketitle

\begin{abstract}
In the analysis of experiments designed to reveal violation of
Bell-type inequalities, it is usually assumed that any hidden
variables associated with the nth particle pair would be independent
of measurement choices and outcomes for the first $(n-1)$ pairs.
Models which violate this assumption exploit what we call the {\it
memory loophole}.  We focus on the strongest type of violation, which
uses the {\it 2-sided} memory loophole, in which the hidden variables
for pair $n$ can depend on the previous measurement choices and
outcomes in both wings of the experiment.  We show that the 2-sided
memory loophole allows a systematic violation of the CHSH inequality
when the data are analysed in the standard way, but cannot produce a
violation if a CHSH expression depending linearly on the data is used.
In the first case, the maximal CHSH violation becomes small as the
number of particle pairs tested becomes large.  Hence, although in
principle the memory loophole implies a slight flaw in existing
analyses of Bell experiments, the data still strongly confirm quantum
mechanics against local hidden variables.

We consider also a related loophole, the {\it simultaneous
measurement loophole}, which applies if all measurements on each side
are carried out simultaneously.  
We show that this can increase the probability of violating the linearised
CHSH inequality as well as other Bell-type inequalities.  

\vskip10pt
PACS number(s): 03.65.-w, 03.65.Ta, 03.65.Ud

\end{abstract}

\vspace{0.2cm}


\section{The memory loophole}

Bell's 
work\cite{bellbook} in the early
1960s made precise the sense in which classical intuitions
based on the principles of special relativity conflict with
quantum theory.  Theoretical and experimental investigations have
continued ever since, leading, inter alia, to the understanding 
of entanglement as a quantifiable resource of fundamental importance
for quantum cryptography, communication and computation.   

The experiment analyzed by Bell is the following \cite{bellbook}. A source
prepares a pair of particles in some entangled state. One particle
is sent to Alice and one to Bob, Alice and Bob being situated far
from each other.  When the particle arrives at Alice, Alice
subjects it to a measurement $X$, chosen by her at random amongst
many possible measurements $A_1$, $A_2$, etc.. Similarly, Bob
subjects his particle to a measurement $Y$ selected by him at
random amongst many possible measurements $B_1$, $B_2$, etc..  The
experiment is repeated many times. Everything is arranged such
that each pair of measurements performed by Alice and Bob is 
space-like separated. After the experiment ends, Alice and Bob
come together and compare their results.

Bell asked whether the correlations between the results of the
measurements predicted by quantum mechanics can be explained by
any classical model. More precisely, he formulated a model, known
as a {\it local hidden variable model}, which is supposed to
describe {\it all possible ways} in which classical systems
can generate correlated answers in an experiment as above. He then
went on to prove that quantum mechanical correlations cannot be
obtained from such a model.

The key words above are ``all possible ways.'' To guarantee that
one has found all possible ways in which a system may behave is a
problematic, and formally not well-defined statement.
Nevertheless, Bell's model, which we describe in detail below, is
very powerful, and it has been generally accepted that it covers
indeed all possibilities. Here however we argue that there are
possibilities that have not been accounted for in Bell's model,
which rely in one way or another on what, for reasons which will be
obvious, we call the {\it memory loophole}.

The rest of our paper is organized as follows. In this section we describe
Bell's original hidden variables model and present the memory
loophole. In section \ref{generalconsiderations}, we introduce an
inequality that is equivalent to the Clauser-Horne-Shimony-Holt (CHSH)
inequality \cite{chsh} and define some terms. In section \ref{thetable}, we
summarize the results of the paper. In sections \ref{nomemory} to
\ref{onesided}, we analyze the inequality from the point of view
of Bell's original model and various different versions of the memory
loophole. We show that the probability of violating a standard
CHSH inequality is affected by the 
loophole, but that the effect is not significant for a large
sample. 
Finally, in section \ref{simultaneous}, we consider a related
loophole which arises in experiments in which all $N$ measurements on
each side are made simultaneously.  
Section \ref{conclusion} concludes.

The model proposed by Bell is the following. When two particles
are prepared at the source in some (entangled) state $\Psi$, they
both receive an index $\lambda$ which is called a {\it local
hidden variable}. This index is chosen at random according to some
distribution $\rho(\lambda)$.  The hidden variable essentially
prescribes, in a {\it local} way, how the particles behave when
subjected to different measurements. That is, when Alice subjects
her particle to a measurement $A$, the particle gives an outcome
$a$ according to some probability distribution $P(a; A,\lambda)$
which depends on the measurement $A$ and on the hidden variable
$\lambda$ but not on the measurement $B$ performed by Bob on his
particle or on the result $b$ of this measurement. Similarly,
Bob's particle yields an outcome $b$ according to the probability
distribution $P(b; B,\lambda)$ which depends on the measurement
$B$ to which it is subjected and on the hidden variable $\lambda$
but not on $A$ or $a$. The joint probability $P(a,b; A,B)$ that
the particles yield  the outcomes $a$ and $b$ when subjected to
the measurement of $A$ and $B$ respectively is then given by

\begin{equation}
P(a, b; A,B)=\int d\lambda\rho(\lambda) P(a; A,
\lambda)P(b;B,\lambda)\label{lhv}
\end{equation}

The above model has been hitherto considered to describe all
possible ways in which classical particles can yield long distance
correlations while respecting the relativistic constraint of no
superluminal signaling, which prevents Alice's particle from modifying
its behavior according to what Bob does if there is not enough
time for a light signal to arrive from Bob to Alice, and
vice versa.   Bell showed that quantum mechanics predicts
correlations which cannot be obtained from such a model.
The inconsistency of quantum theory with the hypothesis of
local hidden variables is often --- slightly confusingly ---
referred to as quantum nonlocality. 

A way of testing whether or not some given correlations can be
obtained from a local hidden variables (LHV) model is to test some
signatures of such models, called Bell inequalities.  The best known
Bell-type inequality is the CHSH inequality.  Suppose that Alice and
Bob chose at random between two measurements $A_1$ or $A_2$ and $B_1$
or $B_2$ respectively. Suppose furthermore that each of these
measurements has only two possible outcomes, $+1$ and $-1$. Then as
CHSH have shown \cite{chsh}, if the particles behave according to any
LHV model,

\begin{equation}
E(A_1 B_1) + E(A_2 B_1) +E(A_1 B_2) - E(A_2B_2)\leq 2,\label{chsh}
\end{equation}
where $E(AB)$ denotes the expectation value of the product of the
outcomes of the measurements $A$ and $B$. On the other hand, one
can find quantum mechanical states $\ket{\Psi}$ (for example, any entangled
pure state \cite{gisinperes,poprohr})
and appropriate measurements so that the CHSH inequality is
violated. For example, if the state $\ket{\Psi}$ is the singlet state of
two spin 1/2 particles,
\begin{equation}
|\Psi\rangle={1\over{\sqrt2}}(|+1\rangle|-1\rangle-|-1\rangle|+1\rangle)
\end{equation}
where $|+1\rangle$ and $|-1\rangle$ represent spin polarised ``up''
and ``down'' along the z axis, one can find appropriate spin
measurements which yield
\begin{equation}
\langle \Psi|A_1B_1|\Psi\rangle + \langle
\Psi|A_2B_1|\Psi\rangle + \langle
\Psi|A_1B_2|\Psi\rangle-\langle \Psi|A_2B_2|\Psi\rangle = 2\sqrt2
\, ,\label{quantumchsh}
\end{equation}
violating the LHV limit (\ref{chsh}). (Here we have used the
quantum mechanical formula for the expectation value
$E(AB)=\langle \Psi|AB|\Psi\rangle$.)

Bell's LHV model has generally been thought to cover all possible
ways in which classical particles can behave. We show now, however,
that this is not the case.

In order to determine correlations one has to perform measurements
not on a single pair of particles but on many such pairs, and
gather a large number of outcomes which will determine the
statistics. Now, according to the LHV model above (\ref{lhv}), all
the pairs in the ensemble are {\it uncorrelated}. This assumption
appears natural from the perspective of quantum mechanics.
In quantum theory, when we have a number of
pairs, each pair being described by the same wave-function, the
pairs are uncorrelated. However, we can imagine the following
scenario. A first pair of particles is emitted by the source. One
of the particles arrives at Alice and it is subjected to a
measurement and gives an outcome according to the LHV model
(\ref{lhv}). However, it also leaves in the environment information
indicating to what measurement it was subjected and what outcome
it yielded. Now, when a particle in the second pair arrives at
Alice, it will read this message and it will give an outcome which
depends not only on the measurement it is subjected to, but also on
the message left by the first particle, i.e. on what has happened
to the first particle. Particles on Bob's side behave in a similar
way. The consequence is that the original LHV model (\ref{lhv}) is
now replaced by
\begin{equation}
P(a^{(n)}, b^{(n)} \, | \, A^{(n)}, B^{(n)})= \int
d\lambda\rho(\lambda) \, 
P(a^{(n)} \, | \, A^{(n)}, M,\lambda) \, P(b^{(n)} \, | \, 
B^{(n)}, M,\lambda) \, ,  \label{newlhv}
\end{equation}
where
\begin{equation}\label{onesideprob} 
P(a^{(n)} \, | \, A^{(n)}, M,\lambda)=
P(a^{(n)} \, | \, A^{(n)},A^{(1)},...,A^{(n-1)},a^{(1)},...,a^{(n-1)}, \lambda)
\end{equation}
and
\begin{equation}
P(b^{(n)} \, | \, B^{(n)}, M,\lambda)=
P(b^{(n)} \, | \, B^{(n)},
B^{(1)},...,B^{(n-1)},b^{(1)},...,b^{(n-1)}, \lambda) \, .
\end{equation}
Here $M$ stands for the local record, or {\it memory}, of the
previous measurements.  We call this a local hidden variable
model with {\it one-sided memory}. 

There is a further interesting variation of Bell's original
model.  Suppose that
the source emits pairs of correlated particles one by one.
Suppose too that on each pair Alice and Bob perform their
measurements space-like separated from one another, so while Alice
is performing her measurement no signal can arrive from Bob's
measurement. However, the time between the measurements on the
different pairs is long enough, so that by the time Alice
measures her $n$-th particle, the particle could have received
information about what has happened in Bob's measurements on all
previous particles ($1,\ldots,n-1$), and similarly for Bob. 
One could imagine local hidden variable models in which this
information is indeed communicated and used, in which case
the probability in (\ref{onesideprob}) is replaced by
\begin{equation}
P(a^{(n)} \, | \, A^{(n)}, M,\lambda)= 
P(a^{(n)} \, | \,  A^{(n)},A^{(1)},...,A^{(n-1)},a^{(1)},...,a^{(n-1)},
  B^{(1)},...,B^{(n-1)},b^{(1)},...,b^{(n-1)}, \lambda)
\end{equation}
and similarly for the probability on Bob's side. 
This is a local hidden variable model with {\it two-sided memory}.   

In principle, Bell's original
argument can be extended to render both types of 
memory loophole irrelevant. 
We could require that separated apparatuses are used for each particle pair,
and that {\it every} measurement is space-like separated from 
every other --- but it seems unlikely that such an experiment
will be done any time soon with a large enough sample of particles
to demonstrate statistically significant violations of Bell inequalities. 
Even the much weaker constraint that all of Alice's 
measurements are space-like separated from all of Bob's --- which
would exclude the two-sided but not the one-sided loophole --- has
not been satisfied in any experiment to date. 
(See, e.g., \cite{aspect} and references therein).  

It is worth emphasizing that the memory loopholes described above have
a different status from that of other loopholes such as the well-known
detection loophole \cite{detectionloophole} or the recently discussed
collapse locality loophole \cite{akloophole}.  The detection loophole
does not identify a problem with Bell's local hidden variables
model {\it per se}, but only states that technological limitations have to
be taken into account --- which can be done in the framework of the
original model.  Similarly, given a precise theory of state reduction
(which is required to characterise the loophole precisely in the first
place), the collapse locality loophole could be closed by carrying out
standard Bell experiments using sufficiently advanced
technology.\cite{akloophole}  On the other hand, although, as
we have noted, the memory loophole could be eliminated by 
new types of Bell experiments, it does highlight an intrinsic 
limitation of Bell's model as applied to standard Bell experiments.  
This is not to
say that local hidden variable theories exploiting the memory loophole
are necessarily more plausible than theories exploiting other
loopholes (indeed, a contrary view can be argued\cite{akloophole}).
It does, though, mean there is need for a reanalysis of the power of
general local hidden variable theories in standard Bell experiments.

Another interesting theoretical question arises if Alice decides to
measure all her particles simultaneously, and Bob does likewise. 
Since the measurements
take some finite time, all the particles on Alice's side could conceivably
communicate to each other.  Hence the outcome given any particle $n$
may depend on what happened with all the other particles (i.e. to
what measurements they are subjected and what outcomes are they
yielding). We call the resulting loophole the
{\it simultaneous measurement loophole} and LHV models 
which exploit it {\it collective} LHV models.  

One might wonder what the point of considering all these loopholes is.
Each seems to involve more conspiracy on Nature's part than the last,
and none of them appears to lead to plausible physical models.  Given
the importance of the Bell-type experiments, however, and their
consequences for our world view, we feel that it is important to
analyze the experiments as rigorously as possible and in particular to
distinguish between logical impossibility and physical implausibility
of the models.  

There is another more practical
motivation\cite{mayersyao,gisinprivcomm}.  It is well known that
quantum key distribution schemes which use entanglement have
significant security advantages over other schemes; they can also be
extended by the use of quantum repeaters to allow secure key
distribution over arbitrary distances.  The security of these schemes
relies crucially on the fact that the states created and measured are
genuinely entangled.  The most obvious and seemingly reliable way to
verify this is to use Bell-type tests as security checks within the
protocols.  However, any such tests need to be interpreted with 
care.   If a quantum cryptosystem is acquired from a not
necessarily reliable source, or possibly exposed to sabotage, then a
cautious user must consider the possibility that devices have been
installed which use classical communication to simulate, as far as
possible, the behaviour of quantum states, while allowing third
parties to extract illicit information about the key.  Such devices
effectively define a local hidden variable model, and the usual
criterion of physical plausibility no longer applies.  A saboteur
could set up communication and computing devices that use any
information available anywhere in the cryptosystem.  In particular,
saboteurs might well try to exploit memory loopholes, as well as 
other Bell experiment loopholes, if they could
gain a significant advantage by so doing.

Having established, therefore, that the original version of the local hidden
variables model as proposed by Bell has to be modified, we
now examine the consequences.

\section{CHSH-type inequalities. General considerations.}\label{generalconsiderations}

We first revisit the usual Bell inequalities experiments,and
emphasize in more detail the statistical aspects of the
measurements.

The standard CHSH inequality is described in (\ref{chsh}) and it
is claimed that every ordinary (i.e. as originally constructed by
Bell) local hidden variables model must obey the inequality.

Of course, even in an ideal experiment, an
ordinary local hidden variables model can violate the
CHSH bound.  The quantities which figure in the CHSH
expression are theoretical expectation values, which are
abstract concepts.  In reality each expectation value is determined
by repeating a measurement a large number of times and estimating
the probabilities (and hence the expectation values) as
frequencies of events.  These measured expectation values
are subject to statistical fluctuations, which can yield violations 
of the CHSH bound.  Our first
task is to examine the problem in detail, defining precisely the
operational meaning of the different quantities, and get an
accurate understanding of what exactly is the meaning of violation
of Bell's inequalities. Only after all these are clarified will we
be able to see the effect of the various memory loopholes. 
In particular, we will see that memory can allow particles to 
take advantage of statistical fluctuations and build them up into a
systematic bias.  We will also see, however, that, if the CHSH
expressions are defined in the usual way, the biases that can
thus be   
obtained tend to zero as the number of pairs tested increases.   
Moreover, we will see that a simpler linearised form of the
CHSH expressions is ``memory-proof'', in the sense that the 
probability of a given level of violation is no greater for
memory-dependent local hidden variable models than for 
optimally chosen memoryless models.   

We use the CHSH inequality in the form \cite{cglmp}
\begin{eqnarray}
P_{CHSH} &=&  P_c(A_1,B_1) + P_c(A_1, B_2) + P_c(A_2,B_1) + 
P_a(A_2,B_2) \nonumber \\ & \leq & 3
\label{probchsh}
\end{eqnarray}
for local hidden variable theories, where $P_c(A,B)$ is the
probability that A and B have the same outcome (are correlated),
and $P_a(A,B)$ is the probability that A and B have different
outcomes (are anti-correlated). $A_1 , A_2 , B_1 , B_2$ are chosen
so that quantum mechanics predicts the maximal value, $P_{CHSH} =
2 + \sqrt{2}$.

What we actually mean by (\ref{probchsh}) in an experimental
context is the following. We suppose that Alice and Bob perform
measurements on $N$ pairs of particles. For each of their
particles Alice and Bob choose at random what measurement to
perform, $A$ or $A'$ for Alice and $B$ or $B'$ for Bob. We define
$\#(A,B)$ to be the number of pairs on which operators A and B
were measured, $\#_c(A,B)$ and $\#_a(A,B)$ to be the number of
times the outcomes were correlated and anti-correlated in these
measurements. Note that Alice and Bob should not pre-arrange the
sequence of their measurements - this would introduce well-known
loopholes; the entire experiments of Alice and Bob, including the
decision of what to measure on each particle have to be space-like
separated from each other. Consequently Alice and Bob do not have
total control on how many times a specific pair of measurements,
say $A,B$ is performed, but this number, $\#(A,B)$ is a random
variable.

We define
\begin{eqnarray}
X_N &=& \frac{\#_c(A_1,B_1)}{\#(A_1,B_1)} +
\frac{\#_c(A_1,B_2)}{\#(A_1,B_2)} + 
\frac{\#_c(A_2,B_1)}{\#(A_2,B_1)} + 
\frac{\#_a(A_2,B_2)}{\#(A_2,B_2)} \, , \label{xdef} \\ 
Y_N &=&
\frac{4}{N} ( \#_c(A_1,B_1) + \#_c(A_1,B_2) + 
\#_c(A_2,B_1) + \#_a(A_2,B_2)) \, .
\end{eqnarray}

$X_N$ is the experimental meaning of the CHSH inequality
(\ref{probchsh}); the index $N$ denotes that the experiment has
been performed on $N$ pairs. Indeed, the expression
$\frac{\#_c(A_1,B_1)}{\#(A_1,B_1)}$ is the frequency of
correlations between the outcomes of $A_1$ and $B_1$, and it is
therefore the experimental definition of the correlation
probability $P_c(A_1,B_1)$ and so on. 

Note that our definition of
$X_N$ assumes that $ \#(A ,B) > 0 $ for all pairs of operators
$A,B$.  If not, $X_N$ is undefined. Strictly speaking, our
expressions for the expectation and other functions of $X_N$
should thus all be conditioned on the event that $X_N$ is defined.
We will neglect this below, assuming that $N$ is large enough that the
probability of $X_N$ being undefined is negligible.  One 
could, alternatively, use an experimental protocol
which ensures that $X_N$ is defined.  For instance, one could
require that, if $ \#(A,B) = 0 $ for any $A,B$ 
after $N$ pairs have been tested, the
experiment continues on further pairs until $ \# (A, B) > 0$ for
all $A,B$, and then terminates.  Our analyses would
need to be modified slightly to apply to such a protocol, 
but the results would be essentially the same.  

$Y_N$ is another experimental quantity closely related to
$X_N$.  The two quantities are equal if the expressions $\# (A_i , B_j )$
are equal for all $i,j$.  For large $N$, the $\# (A_i , B_j )$
are almost always nearly equal, and so the same is true of
$X_N$ and $Y_N$.  Although it is traditional to use 
$X_N$ in analyzing Bell experiments, $Y_N$ is in fact much better
behaved and easier to analyze, since it is a linear expression.

\section{CHSH-type inequalities. Expectation values and 
fluctuations.}\label{thetable}

$X_N$ and $Y_N$ represent quantities determined by making measurements
on a batch of $N$ pairs of particles.  We do not assume the pairs
behave independently: they may be influenced by memory, and we will
analyze the different types of memories.  
We are interested in the maximum possible expectation value of
$X_N$ and $Y_N$, and the maximum probability of $X_N$ or $Y_N$ taking
a value much larger than the expectation.  

Obviously, the expectation and fluctuations of $X_N$ and $Y_N$ could be
experimentally estimated only by repeating 
the whole series of $N$ experiments a large
number of times, and then only under the assumption that 
different batches of $N$ pairs behave independently.  Without
some restriction on the scope of the memory loophole, we would
need to allow for the possibility that any 
experiment we perform in the future could in principle be influenced
by the results obtained in all experiments to date.  

Those who require probabilities to have a frequency interpretation 
in order to be meaningful may thus have some difficulty interpreting
the results of memory loophole analyses.  The only certain way 
to circumvent this difficulty would be to set up many spacelike
separated experiments.     
On the other hand, if probability is viewed simply a measure
of the plausibility of a theory, there is no interpretational
difficulty.  As we will see, it can be 
shown that the probability of obtaining experimental data consistent
with quantum theory, given a local hidden variable theory using 
a memory loophole, for a large sample, is extremely small.  
Since the cumulative data in Bell experiments are indeed 
consistent with quantum theory, we conclude that they 
effectively refute the hypothesis of memory-dependent local hidden 
variables --- so long, of course, as these hidden variables are
assumed not also to exploit other well-known loopholes such as
the detector efficiency loophole.  

The results for which we have complete proofs can be summarized 
in the following table:

\vspace{0.1in}
\begin{tabular}{lcccc}
LHV Model & $E(X_N)$ & $P(\hat{X}_N> 5 \delta)$ & $E(Y_N)$ & $P(\hat{Y}_N>\delta)$ \\[0.8ex]
Memoryless & $ \qquad \leq 3$ & $ < 5f_N^{\delta}$  & $\leq 3$ & $<f_N^{\delta}$ \\
1-sided Memory & $\qquad < 3 + o( N^{-1/2 + \epsilon} ) $
&  $ < 5f_N^{\delta} $ & $\leq 3$ & $<f_N^{\delta}$ \\
Collective & \qquad ?  & ? & $\leq 3 $ & ? \\
2-sided Memory & $\qquad < 3 + o( N^{-1/2 + \epsilon} ) $
& $< 5f_N^{\delta}$ & $\leq 3$ & $<f_N^{\delta}$
\end{tabular}
\vspace{0.1in}

Here $\hat{X}_N = X_N - 3 $, $\hat{Y}_N= Y_N - 3$, and we
have simplified the presentation by taking $\delta$ to be
small enough that $ (3 + \delta ) 
< ( 3 + 5 \delta ) ( 1 - \delta ) $.  
The expression $o (N^{-1/2 + \epsilon})$ denotes a term that 
asymptotically tends to zero faster than $N^{-1/2 + \epsilon}$ for any
$\epsilon > 0$.   
\begin{equation}
f_N^{\delta} = \frac{1}{\sqrt{2 \pi}} \frac{ \sqrt{3}}{
\delta \sqrt{N} } \exp \left( - \frac{1}{6} \delta^2 N  \right).
\end{equation}
The proofs are given in the following sections.

The significance of these results is as follows. The memoryless
case represents the results for standard local hidden variables which
behave independently for each pair.  The
result $E(X_N) \le 3$ is the standard expression of the
CHSH inequality. Although values of $X_N$ larger
than $3$ can be experimentally obtained from a local hidden
variables model, the probability of obtaining $3+\delta$ decreases
exponentially as $5f_N^{\delta}$.  Hence, for a given $\delta$ and
sufficiently large $N$, observing $3+\delta$ 
when performing $N$ experiments can be taken as a very good 
confirmation of the
fact that it is not due to an LHV model.
In the memoryless case, $E(Y_N) \le 3$ and the fluctuations also
decrease exponentially.

In the two-sided memory case, the expectation value of $Y_N$ again satisfies
$E(Y_N) \le 3$. Hence 
the existence of memory makes no difference here.  Memory also
makes no difference to the fluctuations: they still decrease 
exponentially.  On the other hand, the expectation value
of $X_N$ can be larger than in the standard memoryless 
case.   Hypothetically, if Bell experiments are analysed by using
$X_N$ and the effect of the memory loophole is neglected, 
a two-sided memory LHV model could mistakenly be interpreted
as exhibiting non-locality.  
Fortunately, we can put an upper bound of 
\begin{equation} 
E(X_N) \le 3 + 5 N^{-1/2 + \epsilon} + 5 \sqrt{3/ 2 \pi} N^{- \epsilon} 
\exp( - N^{2 \epsilon} / 6 ) \, , 
\end{equation} 
for any small $\epsilon > 0$.   
Thus, for large enough $N$, $X_N$ is almost as 
good  as $Y_N$ at distinguishing quantum theory from local hidden variable
models. 

In the one-sided memory case, we can use the two-sided memory
results to show that $Y_N$ is unaffected by the presence of
memory, and $X_N$ is affected in a negligible way for 
sufficiently large $N$.   Actually, we have not succeeded in
finding a one-sided memory model for which $E(X_N )$ or
$P( \hat{X}_N > \delta )$ are larger than the maximal
values attainable by memoryless models, for any $N$.  
We thus cannot exclude the possibility that
one-sided memory is of no use at all in helping LHV models
come closer to reproducing quantum mechanics. 

In the collective case, $E(Y_N) \le 3$.  However, we 
present
a collective LHV model which has bigger fluctuations
than are possible without memory.  
To have a reliable test of non-locality, we need the 
fluctuations to become small as $N$ gets large.
We conjecture that this is indeed the case: however,
the question marks in the table reflect the fact that we have no
rigorous proof.  

\section{CHSH-type inequalities in Bell's no memory model}\label{nomemory}

We first revisit the derivation of the CHSH inequality in
Bell's model, using techniques which
will be useful for analyzing the different memory models.

We first recall how these quantities are interpreted in standard
analyses, when the Bell pairs are measured sequentially and the
memory loophole is neglected. Let $Z_N$ be 
a binomially distributed variable with $N$ trials, each of 
which has the two possible outcomes $0$ and $1$, with
probability $p \neq 0,1$ of outcome $1$ for each:
The normal approximation to the binomial distribution gives us that 
\begin{equation}
P (Z_N > p N + z \sqrt{N p (1-p)} ) \rightarrow 1 - {\cal N}(z)
\end{equation}
as $N \rightarrow \infty$, where
\begin{equation}
{\cal N}(z) = \frac{1}{\sqrt{2 \pi}} \int_{ - \infty}^z \exp \left( -
\frac{1}{2} y^2 \right) dy
\end{equation}
is the normal distribution function, which obeys
\begin{equation}
1 - {\cal N}(z) \approx \frac{1}{\sqrt{2 \pi}} z^{-1} \exp \left( -
\frac{1}{2} z^2 \right) \, .
\end{equation}
For large $N$, and for $z$ large compared to $1$ and small
compared to $N^{1/2}$, the errors in these approximations are
small and can be rigorously bounded\cite{feller}. Below we
consider $N$ and $z$ in these ranges and neglect the error terms,
which make no essential difference to the discussion.

Now
\begin{equation}
Y_N = \frac{4}{N} \sum_{n=1}^{N} Y_N^n \, ,
\end{equation}
where
\begin{equation}\label{Yterms} Y_N^n = \delta_c^n(A_1,B_1) +
\delta_c^n(A_1,B_2) + \delta_c^n(A_2,B_1) +
\delta_a^n(A_2,B_2) \, .
\end{equation}
Here $\delta_c^n (A,B)$ is $1$ if $A$ and $B$ are measured at the
$n^{th}$ round and found to be the same, and $0$ otherwise, and
$\delta_a^n (A,B)$ is $1$ if $A$ and $B$ are measured at the
$n^{th}$ round and found to be different, and $0$ otherwise.

In a memoryless local hidden variable theory, the $Y_N^n$ are
independent random variables taking values $0$ or $1$. We have
that
\begin{eqnarray}\label{edelta}
E(\delta_c^n(A_1,B_1)) =  \frac{1}{4} p_c^n(A_1 , B_1),
\end{eqnarray}
where $p_c^n(A_1,B_1)$ is the probability that $A_1=B_1$ if
$(A_1,B_1)$ is measured at the $n^{th}$ trial, and similarly for
the other three terms in (\ref{Yterms}). So, from (\ref{chsh}) we
have that
\begin{equation}\label{yexp}
y_n = E(Y_N^n) = \frac{P_{CHSH}}{4} \le \frac{3}{4} \, .
\end{equation}
Clearly, for any $N$ and any $\delta >0 $, the probability $P (
Y_N > 3 + \delta )$ is maximised when the $Y^n_N$ are identically
distributed, with $y_n =3/4$ for all $n$. For small $\delta$ we
have that
\begin{eqnarray}
P ( Y_N > 3 + \delta ) &=& P ( N Y_N /4 > 3 N/ 4 + \delta N / 4 )
\nonumber \\ & \approx & 1 - {\cal N} ( \delta \sqrt{N} / \sqrt{3} )
\nonumber \\ & \approx & \frac{1}{\sqrt{2 \pi}} \frac{ \sqrt{3}}{
\delta \sqrt{N} } \exp \left( - \frac{1}{6} \delta^2 N  \right) \label{yprob}
\, ,
\end{eqnarray}
for large $N$, which tends to zero fast as $N \rightarrow \infty$.
A similar argument shows that quantum mechanics predicts that
\mbox{$P( Y_N < 2 + \sqrt{2} - \delta )$} tends to zero fast.  A
long run of experiments can thus distinguish quantum mechanics and
memoryless local hidden variables with near certainty.

Although the analysis of $Y_N$ is simpler and arguably more
natural, Bell experiments are traditionally interpreted via the
quantity $X_N$.    Since
\begin{eqnarray}
E \left( \frac{\#_c(A,B)}{\#(A,B)} \right)  
& = & \sum_{n=1}^{N} p(\#(A,B)=n) \frac{E(\#_c(A,B) | \#(A,B) =
n)}{n} \nonumber \\ & = & \sum_{n=1}^N p(\#(A,B)=n) \frac{n P_c
(A,B)}{n} \nonumber \\ & = & P_c (A,B) \, ,
\end{eqnarray}
and similarly $ E ( \frac{\#_a(A,B)}{\#(A,B)} ) = P_a (A, B)$,
equations (\ref{chsh}) and (\ref{xdef}) imply that $E(X_N) \le 3$.
(Recall that we assume the $n=0$ terms in these sums have
negligible probability.)

Moreover, since
\begin{equation}\label{xdenom}
P ( \#(A,B) < N/4 (1 - \delta ) ) \approx \frac{ \sqrt{3}}{ \delta
\sqrt{2 \pi N} } \exp \left( - \frac{1}{6}\delta^2 N \right) \, ,
\end{equation}
we have that
\begin{equation}\label{xyratio}
P \left( X_N  >  \frac{1}{1 - \delta } Y_N \right) \lesssim \frac{ 4 \sqrt{3}}{
\delta \sqrt{2 \pi N} } \exp \left( - \frac{1}{6} \delta^2 N  \right)
\end{equation}
and
\begin{equation}\label{xbound}
P \left( X_N  >  \frac{3 + \delta}{1 - \delta } \right) \lesssim \frac{ 5
\sqrt{3}}{ \delta \sqrt{2 \pi N} } \exp \left( - \frac{1}{6} \delta^2
N \right)  \, .
\end{equation}
Similarly, quantum mechanics predicts that \mbox{$P( X_N < 2 +
\sqrt{2} - \delta )$} tends to zero fast.
Thus, for large $N$, $X_N$ distinguishes the predictions of
quantum mechanics and memoryless local hidden variables almost as
well as $Y_N$ does.

\section{The two-sided memory loophole}

Now we consider the case where the LHV model for N trials is
allowed to exploit the memory loophole, predicting results at each
round of measurement which may depend upon the previous
measurements and outcomes on both sides.

Since equations (\ref{edelta}) and (\ref{yexp}) still hold, we
have that
\begin{equation}
E(Y_N)  =  \frac{4}{N} \sum_{n=1}^N E(Y_N^n) 
 \le  \frac{4}{N} \sum_{n=1}^N \frac{3}{4}  =  3 \label{expectation} \, .
\end{equation}

Thus memory does not help increase $E(Y_N)$.  We shall now show 
that it does not help the probability of a large fluctuation in
$Y_N$.  First, we note that $Y_N$ is just (a constant times) 
the sum of $Y_N^n$, where $Y_N^n$ is a random variable at 
the n$^{th}$ trial.  Now, $Y_N^n$ can only take values of $0$ or 
$1$.  To maximize the probability of a large $Y_N$, we should 
try to maximize the probability of each $Y_N^n$ being $1$.  This
at first appears complicated, since with memory LHV models there 
will be correlations between $P(Y_N^n=1)$ for different $n$.  The 
key is to note that, regardless of what happens in later rounds,
for all LHV memory models, 
\begin{equation}
P (Y^n_N = 1 \, | \, \mbox{events in trials $1\ldots n-1$}) \leq 3/4 \, .
\end{equation}
This is because, for any fixed set of events in the earlier rounds, 
the model in round $n$ is just an LHV model, whose probabilities 
have been chosen with no prior knowledge of the measurements 
which will be performed in round $n$, and must therefore satisfy 
the CHSH inequality.  

It follows that, for any $N$ and any $\delta >0 $, the probability
$P ( Y_N > 3 + \delta )$ is maximised when $P(Y_N^n = 1) = 3/4$ for all $n$.
But an LHV model can maximize the probability that  $Y_N^n = 1$,
for any $n$, by a strategy independent of the outcomes of the
previous measurements, for instance by predicting the outcome $1$
for any measurement on either side.  Since $Y_N^n = 0$ or $1$, any
such strategy maximizes the probability $P ( Y_N > 3 + \delta )$,
and so equation (\ref{yprob}) still holds even when the memory
loophole is taken into account. The memory loophole does not alter
the distinguishability of the predictions of quantum mechanics and
local hidden variables, if $Y_N$ is used as the correlation
measure, since neither the maximal expectation nor the maximal
variance of $Y_N$ are increased by memory-dependent strategies.

Now let us turn to $X_N$.  We know that if the particles are
described by identical LHV models, then $E(X_N) \le 3$.  Also,
even when the particles have memory, equations
(\ref{xdenom}-\ref{xbound}) hold.
Suppose we take $ \delta = N^{-1/2 + \epsilon}$, for 
some small $\epsilon > 0$, 
and let $N$ be large enough 
that $\frac{3 + \delta}{1 - \delta} < 3 + 5 \delta$.
Then from (\ref{xbound}), since $X_N$ is always bounded by $4$,  
we have that 
\begin{eqnarray}
E ( X_N ) & \leq & 4 P ( X_N > 3 + 5 \delta ) + ( 3 + 5 \delta ) 
( 1 - P ( X_N > 3 + 5 \delta )) \nonumber \\
& \lesssim & 3 + 5 N^{-1/2 + \epsilon} + 5 \sqrt{3/ 2 \pi} N^{- \epsilon} 
\exp( - N^{2 \epsilon} / 6 ) \, , 
\end{eqnarray} 
so that $( E ( X_N ) - 3 )$ is bounded by a term
that decays faster than $N^{-1/2 + \epsilon}$, for any
$\epsilon > 0$. 
This means that no LHV model can
produce $E( X_N )$ much above $3$ for large $N$; it also means
that the $X_N$ remain efficient discriminators of quantum
mechanics and local hidden variable theories even when the memory
loophole is taken into account.

So far we have shown that the memory loophole makes no essential
difference to Bell inequalities, so long as we use a large number
of particles.  We shall now show that if we only use a small 
number of particles, the two-sided memory loophole does indeed make a 
difference.  We shall give a memory-dependent LHV model
with $E(X_N)
> 3$. To construct a simple example, we take a model which
gives $X_N=3$ with certainty, and modify it a little so that the
expectation increases above $3$.  We set $N=101$.  We can get
$X_{101} = 3$, with certainty, simply by outputing $+1$ regardless
of the observables measured.  Our new model is identical to this
one, except for the case when, after $100$ measurements, we have
measured $(A_1,B_1)$, $(A_1 , B_2 )$ and $(A_2 , B_1 )$  $33$
times each, and $(A_2 , B_2 )$ once. Our new model is allowed memory,
so it can count how many times the various observables are
measured, and thus tell when this is the case. In this (rather
unlikely) case, the new model will output $+1$ on side $A$
regardless of the measurement, and $B_1=+1$, if measured, while
$B_2=-1$, if measured.  
 
The two models will give identical values for $X_{101}$ unless the
above unusual state of affairs occurs after $100$ rounds.
Conditioned upon this event occurring, the old model still has an
expectation of $X_{101}$ equal to $3$, whereas the new model has
slightly more, almost $25/8$. Since the expectation of the new
model is $3$ in all other cases, this increases the unconditional
expectation of the new model to very slightly greater than $3$.

The intuition behind the modification is that if one term in $X_N$
(e.g. $\frac{\#_c(A_1,B_1)}{\#(A_1,B_1)}$) has a small denominator compared
to another term, then we will gain more by increasing the 
numerator in the term with the small denominator than in 
the term with the big denominator.

Now that we have this model with $E(X_N) > 3$, it is easy to see
how to modify it to make a model which does better.  The idea
is to start trying to increase the numerator in the best places
from the start. In each round, there are $4$ possible pairs of
observables which could be measured ($(A_1,B_1)$, $(A_1,B_2)$, etc.).
We can send a list which is guaranteed to give the correct sort of
correlation or anticorrelation to at most $3$ of the possible pairs,
where we can choose which ones.  So at each stage our model must
choose one pair which, if measured, will give the wrong sort of
correlation.  After all the measurements are finished, the model 
would like to give the ``incorrect'' correlation to the pair of
observables which has been measured most (since this term has the
biggest denominator).  There is no way for it to be
sure of doing this, since it does not know at the start which pair
will be measured most.  So, our new model simply guesses.

More precisely, the improved model is as follows.  In the first round
of measurements it gives outcome $+1$, whatever is measured.  From the
second round it looks to see which pair, eg. $(A_1,B_2)$, has been
measured most, and arranges that if that pair is measured in the next
round, the correlations will be "incorrect", whereas if any other pair
is measured in the next round the correlations will be "correct".  It
is easy to see this model produces $E(X_N)> 3$ for all $N$ large
enough that there is a negligible probability of one of the four
observable pairs not being measured.  Of course, our earlier bounds imply
that $ E(X_N ) \rightarrow 3$ as $N \rightarrow \infty$.  We
conjecture that the model produces the maximum value of $E(X_N )$
attainable by a local hidden variable theory with two-sided memory.

\section{The One-Sided Memory Loophole}\label{onesided} 

We comment briefly on the case of the 1-sided memory loophole,
represented by a model of the form (\ref{newlhv}).  
We do not know whether such models 
can increase the value of $E(X_N )$ above $3$, or come any closer
to simulating quantum theory than memoryless LHV models.  
Note, however, that 1-sided memory models are a restricted class of 
the two-sided memory models, and thus all the upper bounds proven for
two-sided models still apply.  In particular, $E(Y_N) \le 3$, and 
equation (\ref{yprob}) still holds, ie. 
$P ( Y_N > 3 + \delta ) \approx \frac{1}{\sqrt{2 \pi}} \frac{ \sqrt{3}}{
\delta \sqrt{N} } \exp ( - \frac{1}{6} \delta^2 N  ).$  
These are in fact tight bounds, since they can be obtained without any memory.

The two-sided bounds also apply for $X_N$.  However, we do not know 
whether they are tight: it may be that one-sided memory LHV models are 
no more powerful than memoryless LHV models.       

\section{Simultaneous measurement loophole}\label{simultaneous}

Although we have seen that the memory loophole gives LHV models some
small wiggle room, it makes little essential difference.  Both
$Y_N$ and $X_N$ remain efficient discriminators between the predictions
of quantum mechanics and of LHV models in the presence of the memory
loophole.  One might conjecture that this is also true of the
simultaneous measurement loophole, and that this can be shown
by essentially the same argument.

However, things are more complicated.  It is true that 
$E(Y_N) \le 3$, since equation (\ref{expectation}) still holds,
following the same reasoning as before.  
However, the fluctuations cannot be dealt with so easily.  
Our arguments to date have
relied on the fact that $P( Y_N^n = 1 ) \leq 3/4$ holds true in
any LHV model, not just a priori, but after conditioning on events
up to round $(n-1)$, even when the memory loophole allows the
behaviour of the round $n$ LHVs to depend on the earlier results.
In particular we have that
\begin{equation}\label{independence}
P( Y_N^n = 1 \, | \,  Y_N^{n-1} = i_{n-1} , \ldots , Y^N_1 = i_1 ) 
\leq 3/4 \,  ,
\end{equation}
for any values of $i_1 , \ldots , i_{n-1}$, in any memory-dependent
LHV model.

However, the derivation of equation (\ref{independence}) relies on the fact that 
results from earlier
rounds are necessarily uncorrelated with measurement choices from
later rounds.  This need not be true when the simultaneous measurement
loophole can be exploited, as the following
simple example illustrates. 

Take $N=2$, and consider simultaneously
measured local hidden variables with the following outcome rules: on
side $A$, the outcomes are $(1,1)$ unless the operators measured are
$(A^1, A^2 )$, when the outcomes are $(1,0)$; on side $B$, the outcomes are
$(1,1)$ unless the operators measured are $(B^2 , B^1 )$, when the outcomes
are $(0,1)$.

Here the outcomes and operators are ordered so that, for
example, an outcome $(i,j)$ means that $i$ was obtained on the relevant
particle from the first pair, and $j$ was obtained on the relevant
particle from the second pair.  The pairs themselves are ordered by
some convention: it does not matter which, so long as the ordering is
consistent on each sides.

It is easy to verify that, in this model,
\begin{equation}
P ( Y_2^1 = 1~ {\rm and} ~ Y_2^2 = 1)  = 10/16 \, ,
\end{equation}
whereas equation
(\ref{independence}) would imply
\begin{equation}
P ( Y_2^1 = 1 ~ {\rm and}~ Y_2^2 = 1)  \leq 9/16 \, .
\end{equation}
In other words, the simultaneous memory loophole allows an LHV model
to increase the probability of getting a larger than expected value
for $Y_2$, beyond that attainable by any model in which the $Y_N^n$
are independent random variables.   The arguments of the preceding
sections thus no longer apply.

We conjecture, nonetheless, that the predictions of quantum mechanics
and of local hidden variables using the simultaneous measurement
loophole can be discriminated by $Y_N$ for large $N$.
If so, then in theory the detector efficiency loophole could be
countered by setting up an experiment in which a single pair
of photons ``simulates'' $N$ spin singlet states: i.e., many degrees
of freedom of a single pair of 
photons are entangled, so that the joint state is isomorphic to 
the state of $N$ singlets.  One could then choose random 
measurements on each photon which simulate independent 
measurements on individual photons in the $N$ singlets. 
Ignoring (admittedly somewhat unrealistically) losses in 
the beam-splitters used to set up the measurements, this 
means that results for all $N$ simulated singlets are
obtained whenever the detectors on both sides fire.  
If both detectors are of efficiency $f$, this 
will happen with probability $f^2$ --- a gain of 
$f^{2 N - 2 }$ over the probability of obtaining a full set of
results if $N$ pairs of photons were separately measured.  
Choosing small $\epsilon$, and taking
$N$ such that $P (Y_N > 3 + \epsilon ) \ll f^2 $ for 
any collective local hidden variables model, would allow
the hypothesis of collective local hidden variables to
be refuted in a single successful experiment (which will
take approximately $f^{-2}$ attempts).   

A strategy for combatting detector efficiency which uses the 
same basic idea of working with a highly entangled state of 
two photons, but is conceptually rather different, has been 
proposed by Massar\cite{massar}.  

\section{Conclusion}\label{conclusion}

We have seen that in the analysis of Bell-type experiments, one ought
to allow for the possibility that the particles have memory, in the
sense that outcomes of measurements on the $n$th pair of particles
depend on both measurement choices and outcomes for the $1{\rm
st},\ldots,(n-1){\rm th}$ pairs. The standard form for local hidden
variable models, originally due to Bell and summarized in equation
(\ref{lhv}), does not allow for this possibility, so a new analysis is needed. We have distinguished
one-sided and two-sided versions of this loophole and shown that in the
two-sided case, a systematic violation of a Bell-type inequality can be
obtained.  In the case of
the CHSH inequality, however, we have derived an upper bound on the
probability of large deviations and thereby shown that the expected
violation tends to zero as the number of particle pairs tested becomes
large. Thus the CHSH inequality is robust against the memory loophole
and the corresponding experimental tests remain good discriminators
between quantum mechanics and local hidden variables --- there is no need 
to design improved experiments in which more (or even all) measurements
are space-like separated from one another.

We have also shown that if the analysis is performed in terms of the
quantities $Y_N$, rather than $X_N$, then the memory
models give no advantage over standard, memoryless, local hidden
variables. Finally, we have considered a related loophole, the
simultaneous measurement loophole, which would arise if Alice and Bob
each performed all his measurements simultaneously, thus allowing for
collective local hidden variables. We have seen that
in this case, the probability of a significant deviation above the CHSH 
bound can be larger than would be allowed for a standard local
hidden variable model. However, we suspect that this extra freedom is
small, in the sense that the predictions of collective local hidden
variables can be distinguished from those of quantum mechanics for a
large enough number of particle pairs. 

\vskip5pt

\leftline{\bf Note added}  \qquad  After this work was completed, 
we became aware of independent work by Gill\cite{gill}, in which similar
bounds on the probabilities of simulating quantum mechanical
results via memory loophole local hidden variable models are
presented.  The existence of the memory loophole was independently
noticed by Accardi and Regoli,\cite{accardiregoli} whose speculation
that it might allow local hidden variables to simulate quantum mechanics
is refuted by Gill's (and our) analyses.
\vskip5pt
\leftline{\bf Acknowledgments} \qquad  We thank Nicolas Gisin for
several helpful discussions.  This work was supported by the
European project EQUIP

\end{document}